\documentclass[aps,prl,twocolumn,showpacs,superscriptaddress,groupedaddress]{revtex4}
\usepackage{amsfonts}
\usepackage{amsmath}
\usepackage{amssymb}
\usepackage{graphicx}
\usepackage{epsfig}
\usepackage{subfigure}
\usepackage{booktabs}
\usepackage{natbib}

\begin{document}

\title{Quantum Electrodynamics is Crucial for Plasmonic Resonance of
Metallic Nanostructures}
\author{Mingliang Zhang, Hongping Xiang, Xu Zhang, and Gang Lu$^{*}$}
\affiliation{Department of Physics and Astronomy, California State University Northridge,
Northridge, CA 91330}
\date{June 3, 2015}

\begin{abstract}
Plasmonic resonance of a metallic nanostructure results from
coherent motion of its conduction electrons driven by incident light. At the
resonance, the induced dipole in the nanostructure is proportional to the
number of the conduction electrons, hence $10^{7}$ times larger than that in
an atom. The interaction energy between the induced dipole and
fluctuating virtual field of the incident light can reach a few tenths of an
eV. Therefore, the classical electromagnetism dominating the field becomes inadequate.
We argue that quantum electrodynamics (QED) should be used instead as the fundamental theory to
describe the virtual field and its interaction with the electrons. Based on QED, we derive analytic
expressions for the plasmonic resonant frequency, which depends on three
easily accessible material parameters. The analytic theory reproduces very
well the experimental data, and can be used for rational design of materials.
\end{abstract}
\maketitle

When light interacts with a metallic nanostructure, its conduction electrons
may undergo collective oscillations driven by the electric field of the
light. Known as localized surface plasmon resonance (LSPR), the collective
oscillations can be tuned by adjusting the shape, size and surrounding
medium of the nanostructure, which is at the heart of the burgeoning field
of plasmonics with potential applications ranging from photocatalysis \cite%
{car,lin} to optics, chemical and biological sensing \cite{may}, and
photo-thermal therapeutics, to name a few \cite{hal4,nke5,kal6,stu7,hu8}.

Although the fundamental physical theory of light-matter interaction is
quantum electrodynamics (QED) \cite{tan,v4}, QED has not been brought to
bear on problems in plasmonics. The present research paradigm of
plasmonics is rooted in classical electrodynamics where the electromagnetic
field of the light is treated classically. It is taken for granted that QED
correction is too small to be relevant in practical plasmonic
applications. The tremendous success of the classical theory has certainly
reinforced this notion. However, as we will show in this paper that the QED
correction \textit{cannot} be ignored in plasmonic resonance. In fact, for
collective excitations such as LSPR, the QED correction could result in an
energy shift on the order of a few tenths of an eV, well within the
experimental probes. For nanoparticles, neglecting the correction would lead
to a size-dependence of resonant frequency that is contradictory to
experiments. The main purpose of the paper is to highlight the importance
and elucidate the consequence of QED in plasmonics. Focusing on plasmonic
resonance of metallic nanostructures, we illustrate the origin of the
plasmonic energy shift and derive analytic expressions for the resonant
frequency. The theory is compared to available experimental results on
nano-spheres, nano-rods and nano-plates and shows promise as a rapid means
for screening materials and structures.

\begin{figure}[ph]
\centering
\par
\subfigure[]{\includegraphics[width=0.23\textwidth]{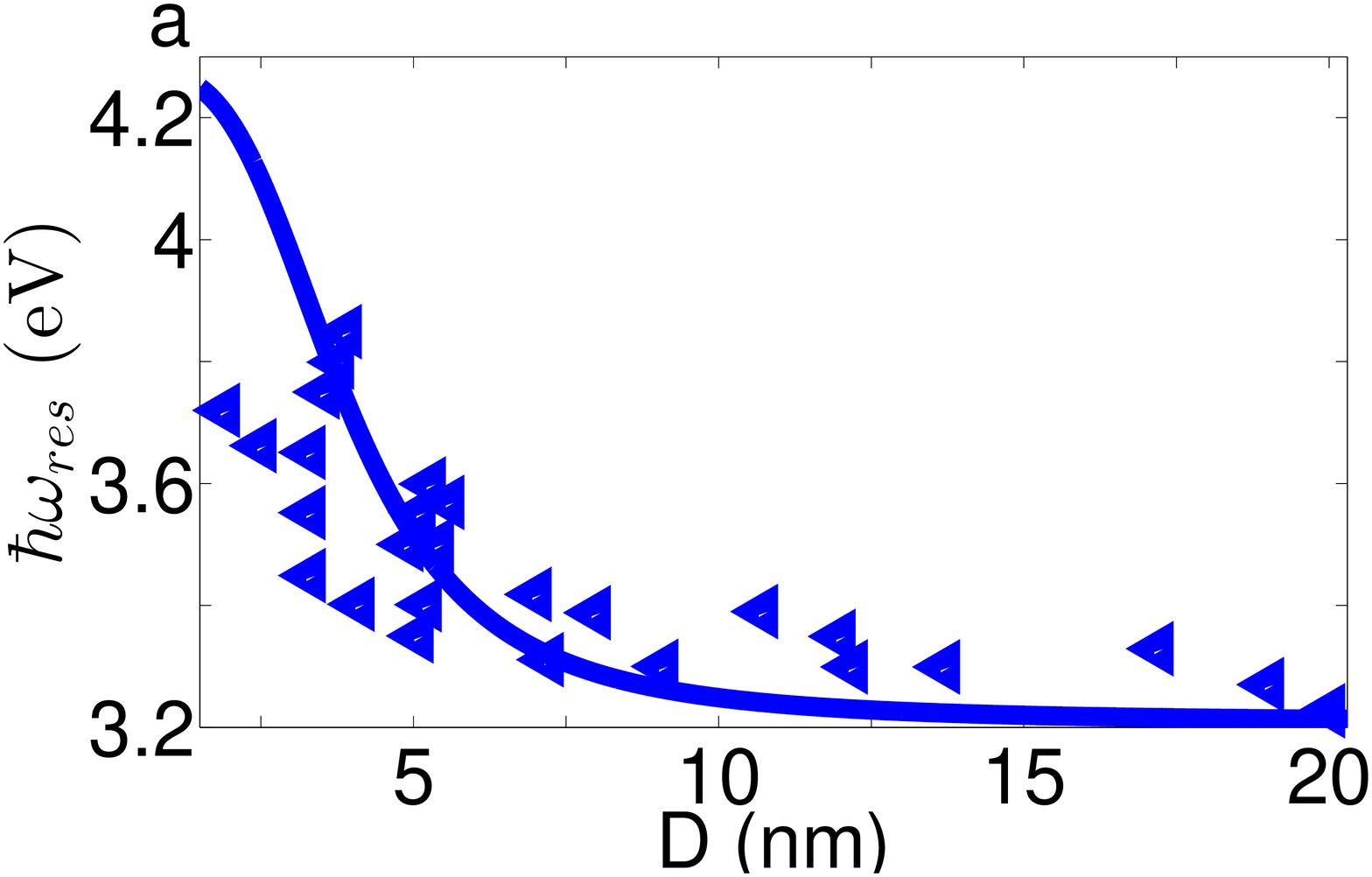}\label{fig1a}} %
\subfigure[]{\includegraphics[width=0.23\textwidth]{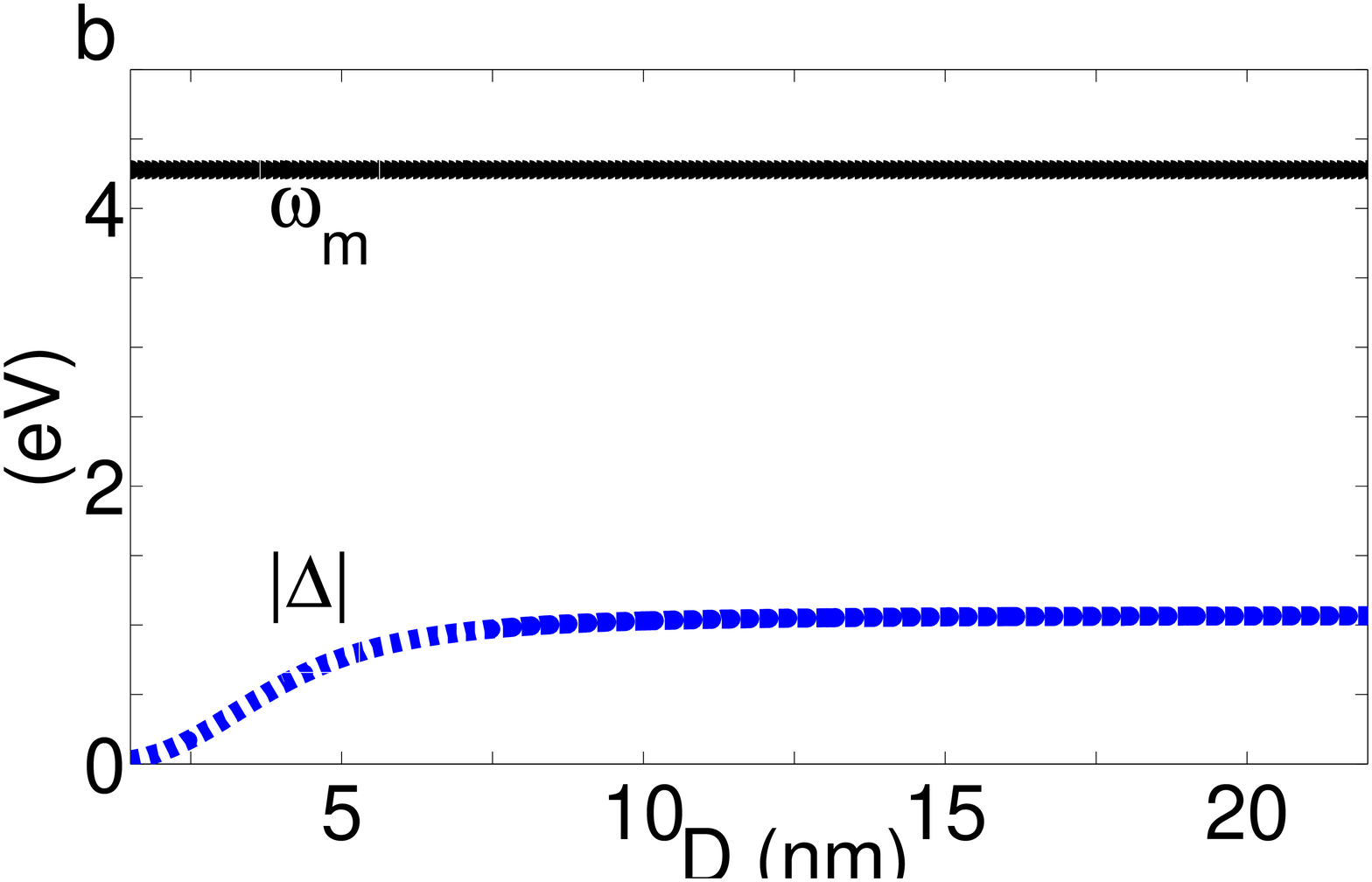}\label{fig1b}}\newline
\caption{(a) The plasmonic resonance energy $\hbar\protect\omega_{\mathrm{res%
}}$ as a function of the nano-sphere diameter $D$, compared between the
theory (solid curve) and experimental data (triangles). (b)The contribution
of $\protect\omega_{\mathrm{m}}$ (dot) and $|\Delta|$ (dash) as a function
of $D$.}
\label{fig1}
\end{figure}

The present theory of plasmonic resonance rests on a classical description
of the electromagnetic field, irrespective of whether the electrons are
treated quantum mechanically or not. In QED, the electromagnetic field is
quantized and there exists a virtual fluctuating electromagnetic field in
vacuum carrying zero-point energy. The fluctuating virtual
field results in a fluctuation of electronic coordinates, leading to a
change in the Coulomb energy between the electrons and the positive ions.
This energy change owing to the virtual electromagnetic field is known as
Lamb shift in atomic physics, and usually extremely small \cite{v4}. In a
plasmonic nanostructure, the wavelength of the visible light, the beam width
and the skin-depth are typically greater than the size
of the nanostructure \cite{cra, v8}, hence {\textit{all} conduction electrons in the
nanostructure undergo coherent and collective oscillations. For a
nanoparticle of a radius of 100 nm, the coherent oscillation could involve $%
\sim10^{7}$ conduction electrons, leading to an induced dipole moment that
is 6-7 orders of magnitude stronger than that in an atom. Since the
interaction energy of the fluctuating virtual field with the conduction
electrons is proportional to the induced dipole, the energy shift can reach
several tenths of an eV, which can be measured experimentally. }

Let $H_{e}$ be the electronic Hamiltonian of the nanostructure, and $%
\{|a\rangle ,|b\rangle ,\cdots \}$ and $\{E_{a},E_{b},\cdots \}$ as the
eigenstates and eigenvalues of $H_{e}$, respectively. Let $H_{f}$ be the
Hamiltonian of the incident electromagnetic field including both the
external field and the virtual field. The eigenstates of $H_{f}$
are labeled as $|n_{\mathbf{k}_{1}\mathbf{e}_{1}},n_{\mathbf{k}_{2}\mathbf{e}%
_{2}}\cdots \rangle $, where $n_{\mathbf{k}_{1}\mathbf{e}_{1}}$ is the
occupation number in the photon state $\mathbf{k}_{1}\mathbf{e}_{1}$ ($%
\mathbf{k}_{1}$ is the wave-vector and $\mathbf{e}_{1}$ is the polarization
vector). The total Hamiltonian of the system is thus
\begin{equation}
H=(H_{e}+H_{f})+U,  \label{ht}
\end{equation}%
where $U$ is the interaction between the conduction electrons and the
virtual field. Since the wavelength of the field is much larger than the
size of the nanostructure, the interaction energy $U$ can be expressed as
\cite{tan,v4}
\begin{equation}
U=-\sum_{j}\widehat{\mathbf{d}}_{j}\cdot \mathbf{E}(\mathbf{r}_{j}),
\label{u}
\end{equation}%
where $\mathbf{E}(\mathbf{r}_{j})$ is the electric field at position $%
\mathbf{r}_{j}$ of the $jth$ electron, and $\widehat{\mathbf{d}}_{j}$ is the
dipole operator of the $jth$ electron. The extinction spectrum of the
nanostructure, comprised of the absorption and scattering spectrum of the
incident photon, is the primary physical quantity that can be measured
experimentally and calculated theoretically. According to QED \cite{tan},
the resonant frequency $\omega _{\mathrm{res}}$ of the extinction spectrum
consists of two contributions:
\begin{equation}
\omega _{\mathrm{res}}=\omega _{m}+\Delta.  \label{res}
\end{equation}%
$\hbar \omega _{m}$ represents the excitation energy of the plasmons. Since
the plasmons or the coherent oscillations are the eigenstates of $H_{e}$,
denoted by $\Psi _{m}$, $\omega _{\mathrm{m}}$ is the corresponding
eigenfrequency. $\Delta $ is the frequency shift resulted from the
interaction of the electrons with the virtual field.

\begin{figure}[ph]
\centering
\subfigure[]{\includegraphics[width=0.23\textwidth]{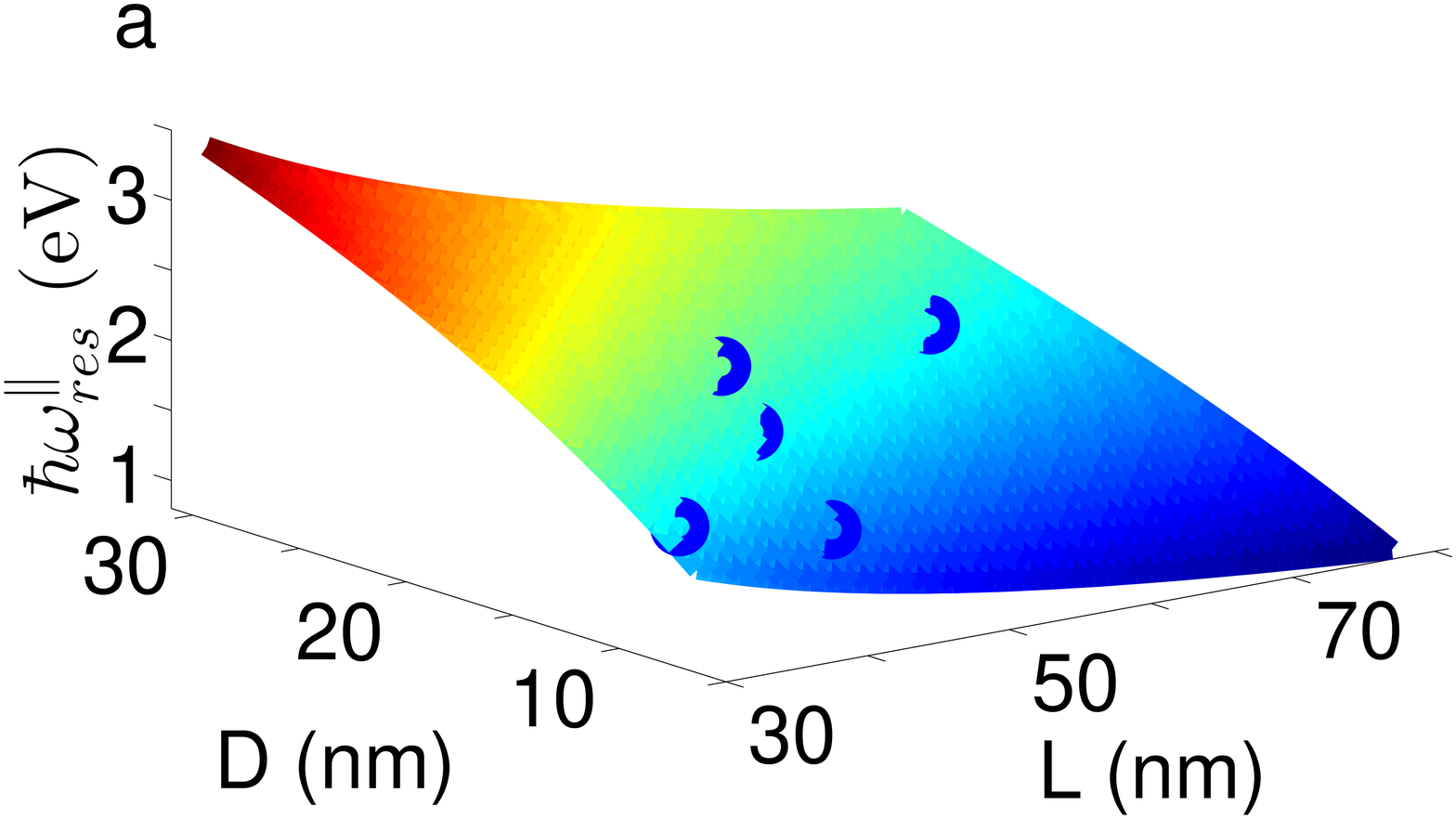}\label{fig2a}}
\subfigure[]{\includegraphics[width=0.23\textwidth]{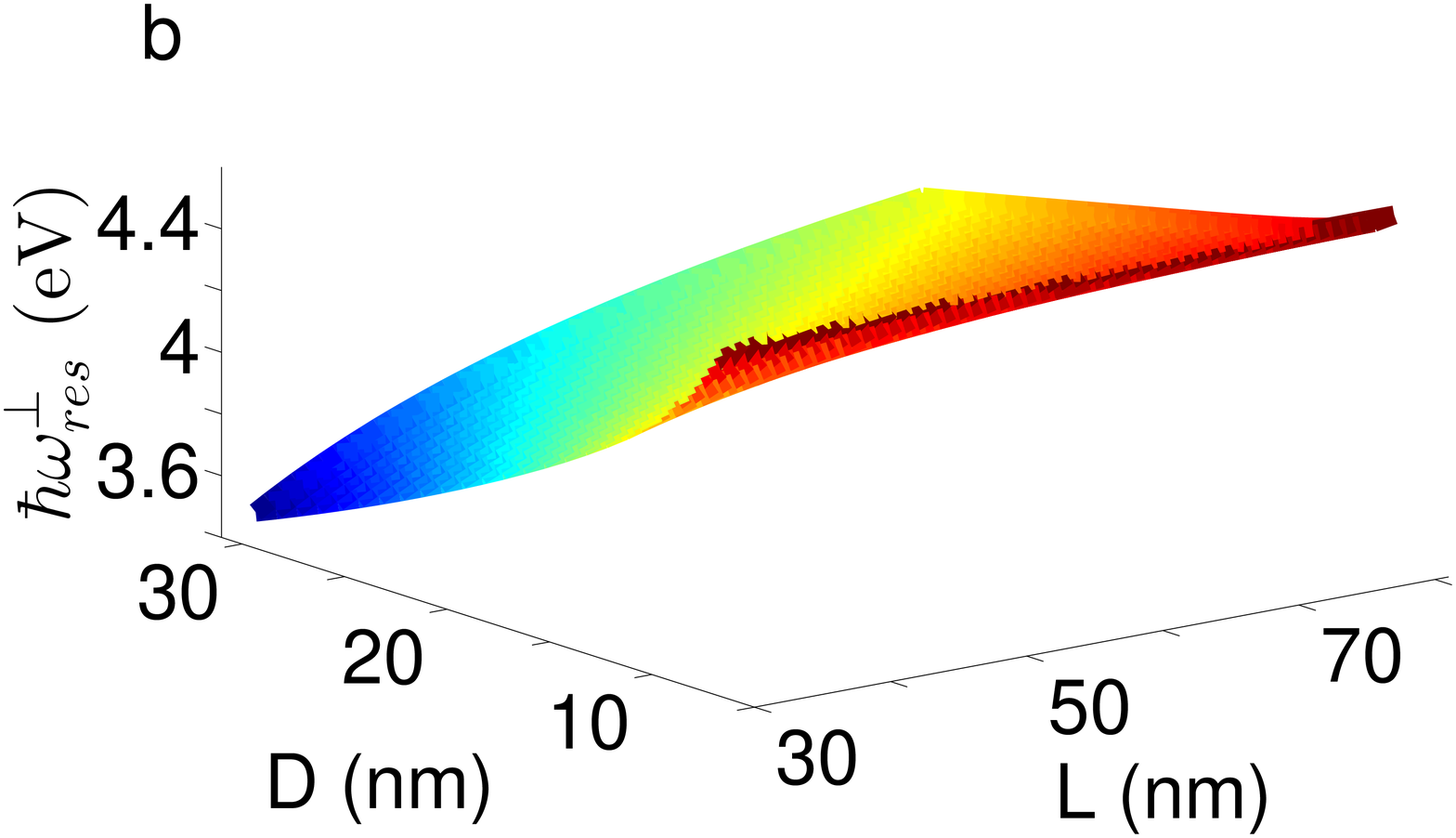}\label{fig2b}}
\caption{The plasmonic resonance energy of a Au nano-rod embedded in SiO$%
_{2} $ as a function of its diameter $D$ and length $L$, compared between
the theory (surface) and experiment (circle). The polarization direction is
parallel (a) and perpendicular (b) to the nano-rod long-axis.}
\label{Aurod}
\end{figure}

Here we derive an analytical expression for this eigenfrequency, with the
oscillations along one of the major axes of the nanostructure. To a good
approximation, nanostructures such as spheres, rods, and circular plates can
be modeled as ellipsoids with a uniform volume polarization. Suppose that a
metallic ellipsoid with a dielectric function $\varepsilon _{1}=\varepsilon
_{1}^{\prime }+i\varepsilon _{1}^{\prime \prime }$ is embedded in a medium
with a dielectric function $\varepsilon _{2}=\varepsilon _{2}^{\prime
}+i\varepsilon _{2}^{\prime \prime }$, and the electric field of the
incident photon, $E_{z}^{\mathrm{ext}}$, is along the z-axis of the
ellipsoid. If the longest dimension the ellipsoid, $L_{\max }$, is much
smaller than the wavelength $\lambda $, the beam width $W$ of the incident
light \cite{cra}, and the skin-depth $\delta =\lambda (2\pi \varepsilon
_{1}^{\prime \prime 1/2})^{-1}$, the total induced dipole moment in the
ellipsoid is given by \cite{str,v8}:
\begin{equation}
\mathcal{P}_{1z}=\alpha _{z}V\epsilon _{0}E_{z}^{\mathrm{ext}},~\mathrm{with}%
~\alpha _{z}=\frac{\varepsilon _{2}(\varepsilon _{1}-1)}{\varepsilon
_{2}+(\varepsilon _{1}-\varepsilon _{2})n^{(z)}}.  \label{dip}
\end{equation}%
Here $V$ is the volume of the ellipsoid; $\epsilon _{0}$ is the permeability
of free space and $n^{(z)}$ is the depolarization factor along the z-axis.
We now make two assumptions: (1) $\varepsilon _{2}^{\prime }$ is a
constant, which is a reasonable assumption for commonly used media, such as vacuum,
SiO$_{2}$ and polyvinyl alcohol, in plasmonic applications. (2)
The dielectric functions of the nanostructures are primarily determined by
the conduction electrons \cite{cal}, which is also a reasonable
approximation. Under these assumptions, the dielectric functions of the
nanostructure can be expressed as:
\begin{equation}
\varepsilon _{1}^{\prime }=1-\frac{S\omega _{p}^{2}}{\omega ^{2}+\gamma ^{2}}%
,~~\varepsilon _{1}^{\prime \prime }=\frac{S\gamma \omega _{p}^{2}}{\omega
^{3}+\gamma ^{2}\omega }.  \label{di1}
\end{equation}%
Here $\omega _{p}=(ne^{2}/m\epsilon _{0})^{1/2}$ is the plasmon frequency. $%
n $ and $m$ denote the electron number density and mass of the electron. $%
S(\omega )$ represents intra-band oscillator strength of the conduction
electrons. $\gamma $ is the decay rate of the quasiparticles, given by \cite%
{cal,lee,co,ja}:
\begin{equation}
\gamma =\gamma _{0}+Av_{\mathrm{F}}/L_{\mathrm{eff}},  \label{bd}
\end{equation}%
where $v_{\mathrm{F}}=\hbar (3\pi ^{2}n)^{1/3}/m$ is the Fermi velocity of
the metal and $L_{\mathrm{eff}}$ is the effective dimension of the ellipsoid
along the polarization direction. $A$ is a dimensionless, positive constant
on the order of one. $\gamma _{0}$ is the decay rate resulting from
electron-phonon, electron-impurity, and electron-electron interactions \cite%
{cal}, and can be estimated by $\gamma _{0}=ne^{2}/m\sigma $ with $\sigma $
as the conductivity of the metal. The second term in Eq.(\ref{bd}) stems
from electron scattering with the surface, and is the only term dependent on
the size of the nanostructure.

\begin{figure}[ph]
\centering
\subfigure[]{\includegraphics[width=0.23\textwidth]{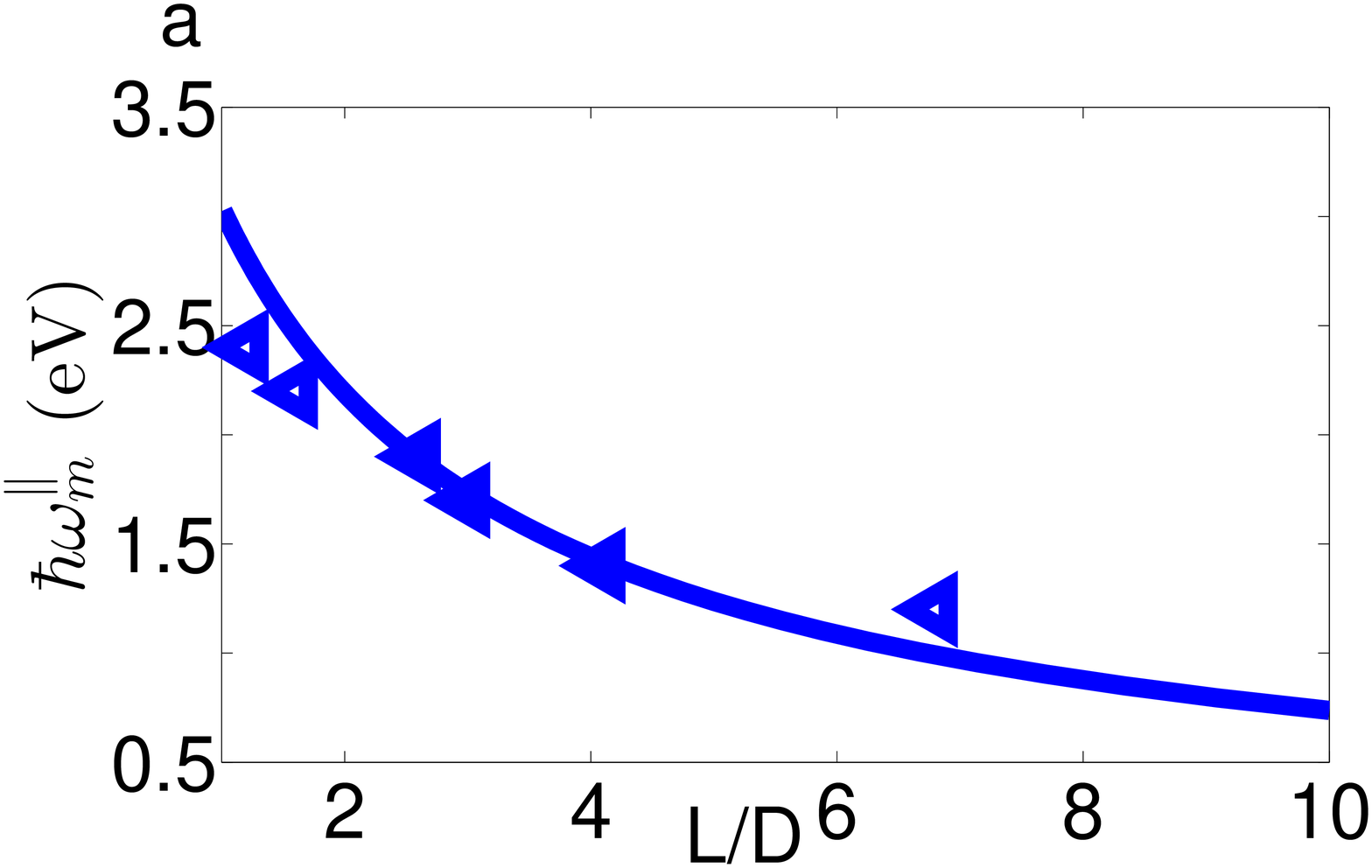}\label{fig3a}}
\subfigure[]{\includegraphics[width=0.23\textwidth]{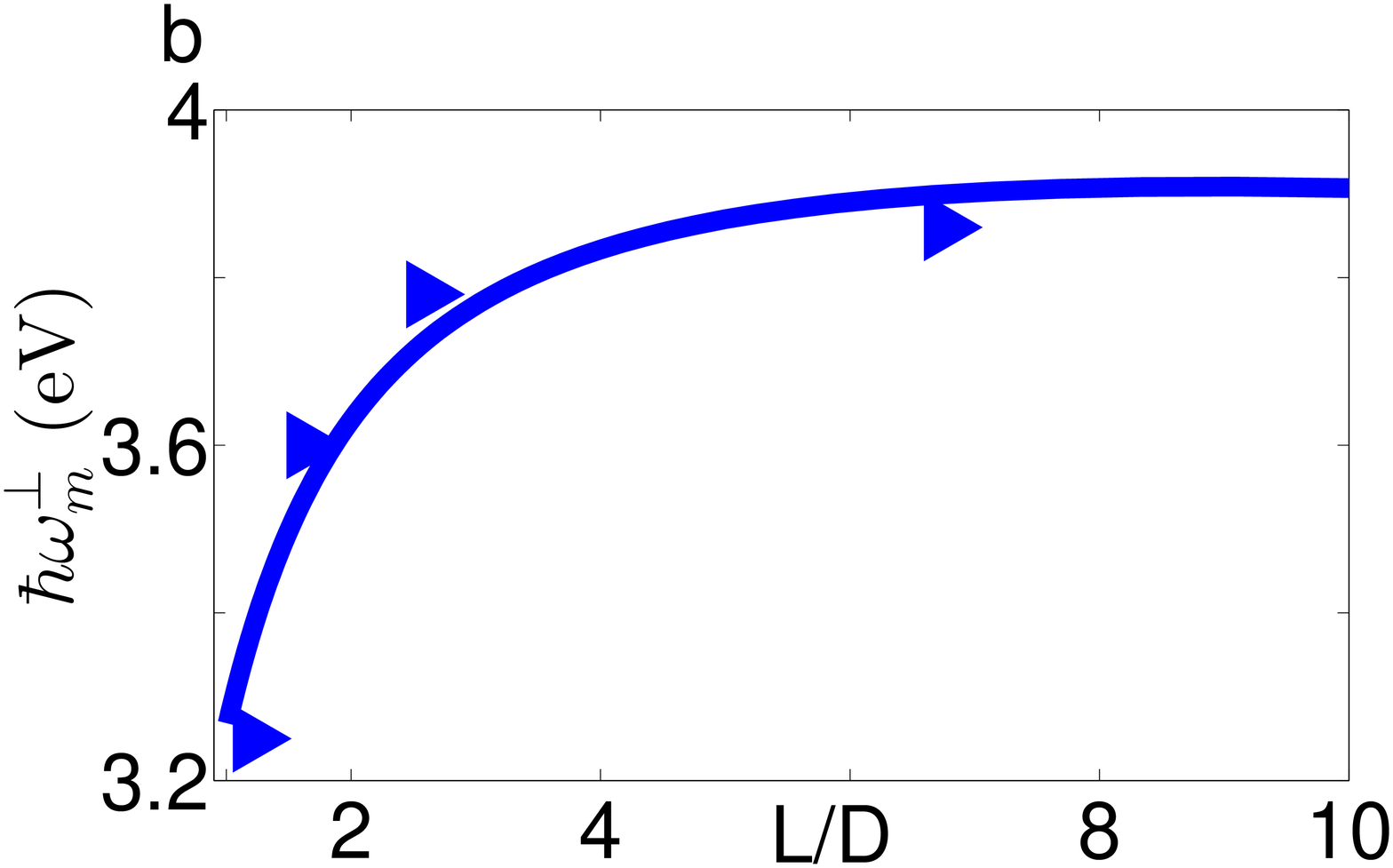}\label{fig3b}}
\caption{The eigenfrequency $\protect\omega_{\mathrm{m}}$ as a function of $%
L/D$ for a Na nano-rod in vacuum, determined from the theory (curve) and the
TD-OFDFT calculations (triangles). The polarization direction is parallel
(a) and perpendicular (b) to the nano-rod long-axis.}
\label{Narod}
\end{figure}

In the classical electromagnetism, the extinction cross-section $\sigma _{%
\mathrm{ex}}^{c}(\omega )$ of the nanostructure is proportional to $%
\{[\varepsilon _{2}^{\prime }+(\varepsilon _{1}^{\prime }-\varepsilon
_{2}^{\prime })n^{(z)}]^{2}+[\varepsilon _{2}^{\prime \prime }+(\varepsilon
_{1}^{\prime \prime }-\varepsilon _{2}^{\prime \prime })n^{(z)}]^{2}\}^{-1}$
\cite{v8}. We show in the Supplementary Information that the eigenfrequency $\omega _{m}$
maximizes $\sigma _{\mathrm{ex}}^{c}(\omega )$.
Therefore $\omega _{m}$ is the root of the following equation \cite%
{zor,ju,ba,sch,ka,vo,krg,ste,ber}:
\begin{equation}
\varepsilon _{2}^{\prime }+(\varepsilon _{1}^{\prime }-\varepsilon
_{2}^{\prime })n^{(z)}=0.  \label{mac}
\end{equation}%
Eqn.(\ref{mac}) yields the expression for the eigenfrequency $\omega _{m}$:
\begin{equation}
\omega _{m}=\omega _{p}[\frac{S}{1+\varepsilon _{2}^{\prime }(\frac{1}{n}-1)}%
-(\frac{\gamma }{\omega _{p}})^{2}]^{1/2}.  \label{eif}
\end{equation}%
Here and later we drop the superscript $z$ for brevity when there is no
confusion.

Based on non-perturbative many-body quantum theory, the frequency shift $%
\Delta $ can be determined from the Green's function of the Hamiltonian (\ref%
{ht}). Specifically, $\hbar \Delta ={\rm Re}(R)$ and $R$ is the
shift-operator of the Green's function, given by \cite{tan}
\begin{equation}
R=\sum_{a\mathbf{ke}}\frac{|\langle a;\mathbf{ke}|U|\Psi _{m};0\rangle |^{2}%
}{E_{m}-E_{a}-\hbar \omega _{\mathbf{ke}}}+\sum_{a\mathbf{ke}%
}\sum_{a^{\prime }\mathbf{k}^{\prime }\mathbf{e}^{\prime }}  \label{sv}
\end{equation}%
\begin{equation*}
\frac{\langle \Psi _{m};0|U|a;\mathbf{ke}\rangle \langle a;\mathbf{ke}%
|U|a^{\prime };\mathbf{k}^{\prime }\mathbf{e}^{\prime }\rangle \langle
a^{\prime };\mathbf{k}^{\prime }\mathbf{e}^{\prime }|U|\Psi _{m};0\rangle }{%
(E_{m}-E_{a}-\hbar \omega _{\mathbf{ke}})(E_{m}-E_{a^{\prime }}-\hbar \omega
_{\mathbf{k}^{\prime }\mathbf{e}^{\prime }})}+\cdots ,
\end{equation*}%
where $E_{m}$ is the energy of the eigenstate $\Psi _{m}$; $|\Psi
_{m};0\rangle $ represents the many-body state at which the incident photon
is absorbed and the plasmon is excited. $|a;\mathbf{ke}\rangle $ denotes
direct-product state of a many-electron state $|a\rangle $ and a virtual
photon state with a wave-vector $\mathbf{k}$ and a polarization vector $%
\mathbf{e}$. Finally, we arrive at
\begin{equation}
\Delta =-\frac{\omega _{m}}{4}\frac{\beta _{z}}{1+\beta _{z}},  \label{sh}
\end{equation}%
with
\begin{equation}
\beta _{z}=\frac{V}{\lambda _{m}^{3}}\frac{\varepsilon _{2}^{\prime 2}}{%
[n^{(z)}]^{2}}\{1+(\frac{\varepsilon _{1}^{\prime }-1}{\varepsilon
_{1}^{\prime \prime }})^{2}\}.  \label{sh1}
\end{equation}%
Here $\lambda _{m}=2\pi c/\omega _{m}$. In Eqn. (\ref{sh1}), $\varepsilon
_{1}^{\prime }$ and $\varepsilon _{1}^{\prime \prime }$ are evaluated at $%
\omega =\omega _{m}$. As $\beta _{z}\geq 0$, the frequency shift $\Delta $
is less than 1/4 of $\omega _{m}$.

According to Eqn.(\ref{eif}, \ref{sh}), there are three contributions to $%
\omega _{\mathrm{res}}$. Among them, the dominant one is the first term in
Eqn. (\ref{eif}) since $\frac{\gamma }{\omega _{p}}<<1$; the second term in
Eqn. (\ref{eif}) is the smallest among them. The dominant term depends \textit{%
only} on the shape of the nanostructure through the depolarization factor, $%
n^{(z)}$. Hence the resonant frequency $\omega _{\mathrm{res}}$ of the
nanostructure is determined primarily by its shape as opposed to its size.
This fact has been well established and exploited in plasmonics \cite{ju,zor}%
. More importantly, $\gamma $ in Eqn.(\ref{eif}) is a monotonically
decreasing function of $L_{\mathrm{eff}}$ as indicated in Eqn.(\ref{bd}),
hence $\omega _{m}$ is a weakly increasing function of the particle size. If
there were no correction term $\Delta $, the resonant frequency $\omega _{%
\mathrm{res}}$ (=$\omega _{m}$) would have been a monotonically increasing
function of the particle size, which is opposite to the experimental
observations \cite{ba,sch,ste,ja}. The QED correction term, $\Delta $, as a
decreasing function of the particle size, reverses the incorrect
size-dependence of the classical electromagnetic theory and renders $\omega
_{\mathrm{res}}$ consistent with the experiments. The failure of the
classical theory has also been discussed by Scholl et al. \cite{sch} who
attributed the opposite size-dependence to the inappropriate use of
macroscopic dielectric functions in the nanoparticles. The macroscopic
dielectric functions failed to capture the effects of discrete energy levels
and the fact that only certain electronic or plasmonic transitions are
allowed in the nanoparticles. To remedy the classical theory, Scholl et al.
proposed a phenomenological model based on discrete energy levels. Although
the model yielded an improved agreement to the experimental data, it did not
consider the quantum effect of the electromagnetic field. As a result, the
model cannot guarantee the \textit{monotonically} decreasing size-dependence
of $\omega _{\mathrm{res}}$, as revealed in experiments and the present
theory. Nonetheless, Scholl's model is valuable contribution and could be
combined with the present theory to form a more comprehensive microscopic
picture of plasmonic resonance.

If the electric field of the incident light is perpendicular to the z-axis,
the depolarization factor in the normal direction has to be worked out. We
have derived the corresponding equations, which are included in the
Supplementary Information. Moreover, if the electric field of the incident
light is along an arbitrary direction, the total induced dipole is a vector
sum of the components in each major axis \cite{str,v8}. Finally, for a
spheroid with its rotational axis along z, the depolarization factors $%
n^{(x)}$, $n^{(y)}$, $n^{(z)}$ and the resonant frequency $\omega _{\mathrm{%
res}}^{\parallel }$ ($\mathbf{E}_{\mathrm{ext}}\parallel $z) and $\omega _{%
\mathrm{res}}^{\perp }$ ($\mathbf{E}_{\mathrm{ext}}\perp $z) can be
calculated analytically as well. For a general ellipsoid, the corresponding
quantities have to be evaluated numerically.

To validate the proposed theory, we apply it to various metallic
nanostructures including nano-spheres, nano-rods, and nano-plates. First, we
examine the size-dependence of $\omega _{\mathrm{res}}$ in nano-spheres.
Since all spheres have the same shape or the depolarization factors ($n=1/3$%
) \cite{v8}, the nano-spheres of the same metal would yield the same $\omega
_{m}$ for a given surrounding medium. According to Eqn.(\ref{sh}, \ref{sh1}%
), $\Delta $ depends only on the volume of a sphere, thus $\omega _{\mathrm{%
res}}$ is a monotonically decreasing function of the sphere diameter $D$. In
Fig. 1(a), we compare the theoretical prediction to the experimental data
taken from Fig. 3(b) of reference \cite{sch} for Ag nano-spheres. The
dielectric constant of the surrounding medium $\varepsilon _{2}^{\prime }$
is 1.69 as measured in the experiment \cite{sch}. We find that the
theoretical prediction agrees very well to the experimental data as long as
the two fitting parameters $A$ and $S$ are chosen reasonably, in this case $%
A=0.03,~S=1$. As displayed in Fig. 1(b), the size dependence of $\omega _{%
\mathrm{res}}$ is entirely contained in $\Delta $ while $\omega _{m}$ is
essentially flat. Thus the size-dependence of the nano-spheres originates
exclusively from the quantum nature of the electromagnetic field.

Second, we compare the theoretical prediction to the experimental results
\cite{ju} for Au nano-rods embedded in silica ($\varepsilon _{2}^{\prime
}=2.15$). In Fig. 2 (a), the experimental resonance frequency $\omega _{%
\mathrm{res}}^{\parallel }$ as function of the length $L$ and diameter $D$
of the nano-rods is shown in circles, while the theoretical prediction is
the surface. In this case, the two fitting parameters are $A=0.6,~S=1.4$.
Because $L\sim$ 32-70 nm \cite{ju}, larger than the size of the
nano-spheres, the electron scattering at the surface becomes more important,
thus $A$ is larger. For the similar reason, the oscillator strength $S$ is
also larger than the nano-spheres, as discussed in the Supplementary
Information. There is an overall good agreement between the theory and
experiment, down to the size of 8.5 nm \cite{ju}. The theoretical prediction
for the polarization direction perpendicular to the nano-rod axis is
displayed in Fig. 2 (b).

We next demonstrate the validity of $\omega _{m}$, which cannot be measured
directly by experiments. Hence, we compare the theoretical prediction of $%
\omega _{m}$ to a set of computational results obtained from time-dependent
orbital-free density functional theory (TD-OFDFT) simulations \cite{hpx} for
a Na nano-rod embedded in vacuum ($\varepsilon _{2}^{\prime }$=1). A number
of (L,D) combinations including (5.79,0.86), (5.79,1.41), (5.79,1.93),
(5.79,2.23), (5.79,3.54), (5.79,4.76), and (5.79,5.46), in the unit of nm,
are considered. The two fitting parameters are $A=0.6,~S=0.8$ for $\mathbf{E}%
\parallel $ axis and $A=0.6$, $S=0.92$ for $\mathbf{E}\perp $ axis. There is
an excellent agreement between the theoretical predictions and the TD-OFDFT
results for both polarization directions as shown in Fig. 3, which validates
the derivation of $\omega _{m}$. In Fig. 2 and 3, one may notice that $\hbar
\omega _{\mathrm{res}}^{\parallel }(L,D)$ has an opposite size-dependence as
$\hbar \omega _{\mathrm{res}}^{\perp }(L,D)$, owing to the opposite (L,D)
dependence of $n^{(z)}$ and $n^{(x)}$.

\begin{figure}[ph]
\centering
\subfigure[]{\includegraphics[width=0.23\textwidth]{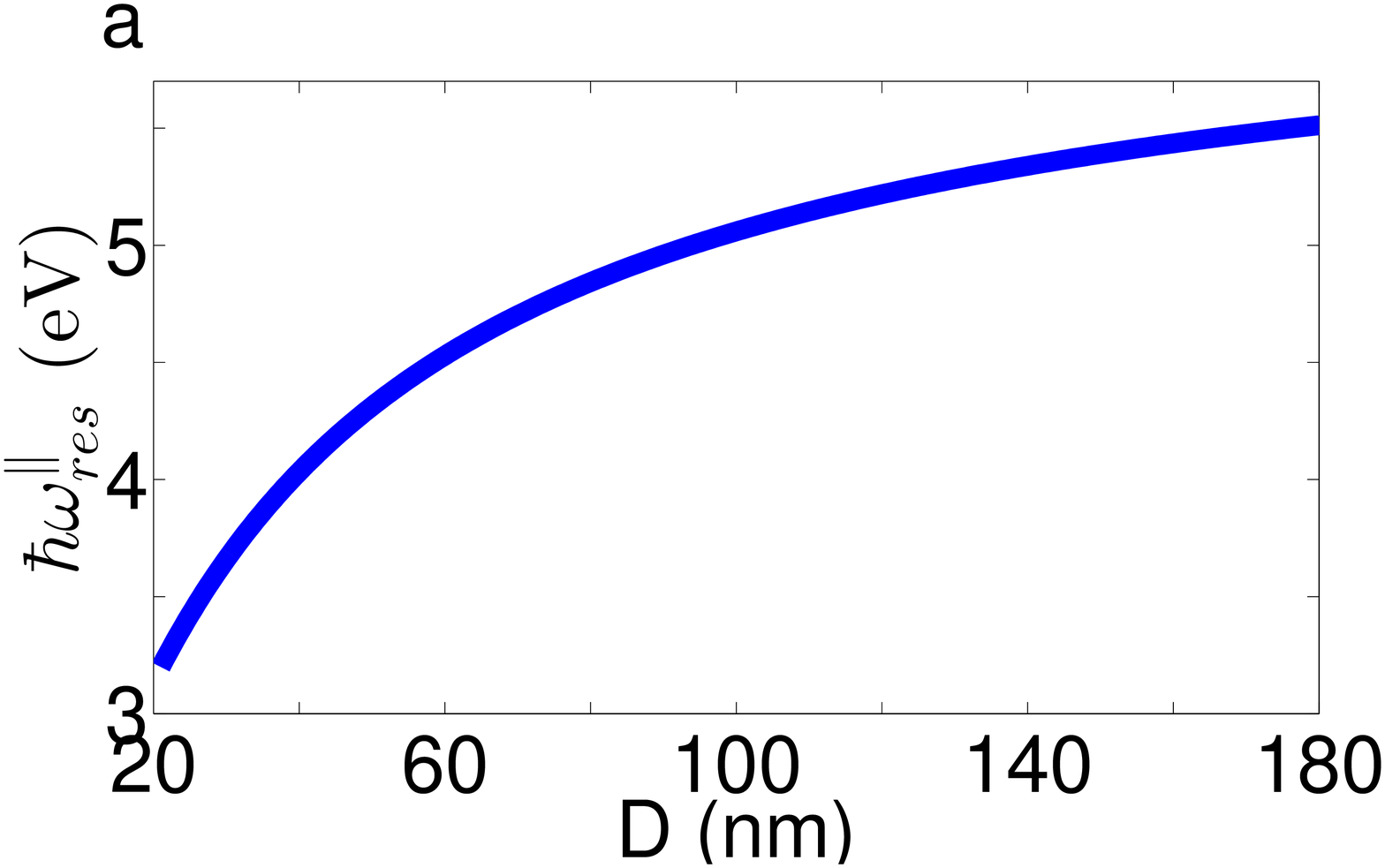}\label{fig4a}}
\subfigure[]{\includegraphics[width=0.23\textwidth]{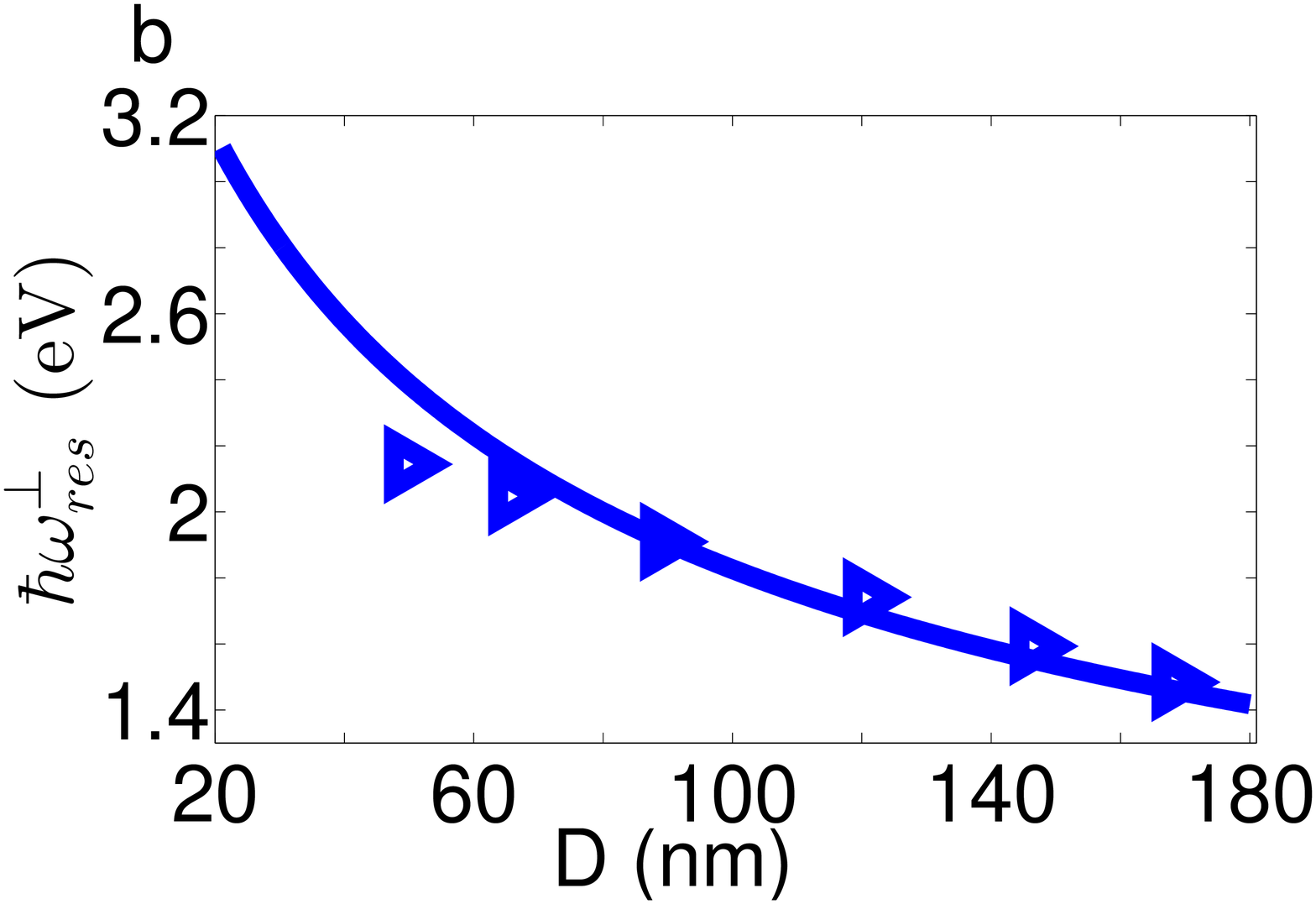}\label{fig4b}}
\caption{The plasmonic resonance energy as a function of diameter $D$ for a
circular gold nano-plate with a fixed height $L$ = 20 nm. The results from
theory are shown in solid curves while the experimental data is shown as
triangles; The polarization direction is parallel (a) and perpendicular (b)
to the long-axis of the plate.}
\label{fig4}
\end{figure}

Third, we switch to a Au nano-plate whose rotational symmetric axis is along
z. The theoretical predictions for $\omega _{\mathrm{res}}^{\perp }(D)$ are
compared to the experimental results \cite{zor}. The circular nano-plate has
a height $L=20$nm embedded in a medium with a refractive index of 1.26. The
fitting parameters $A=0.6$,$~S=0.86$ yield an excellent agreement between
the theoretical predictions and the experimental data as shown in Fig. 4.
Similar agreement between the theory and experiments is also observed for Al
and Pt nano-plates which is presented in the Supplementary Information. The
opposite D-dependence between $\omega _{\mathrm{res}}^{\parallel }(D)$ and $%
\omega _{\mathrm{res}}^{\perp }(D)$ is due to the opposite D-dependence of
their respective depolarization factors.

QED could have more profound implications in plasmonics than what is
presented in this paper. For example, it is known that when a nano-antenna
is placed next to a metallic nanostructure, there is an interaction between
the nano-antenna (an emitter) and the virtual field. The interaction could
change the directional radiation pattern of the antenna \cite{gia},
analogous to cavity QED \cite{har}. The present work, on the other hand,
focuses on the interaction between an absorber (the plasmonic nanostructure)
and the virtual field. Such interaction could also change the induced
magnetic moment of the metallic nanostructure, as well as the polarization
of the incident light.

To summarize, we propose that QED is crucial to understand the plasmonic
resonance in metallic nanostructures. The coherent motion of the conduction electrons
in the nanostructure could lead to a large induced dipole moment, which
interacts with the virtual field and results in a significant shift in the
resonant frequency. The frequency shift is the key to reconciling the
theoretical predictions and experimental observations on size-dependent
plasmonic resonance. Based on QED, we have derived analytic expressions for
the plasmonic resonant frequency, which depends on three easily accessible
material parameters - the dielectric constant $\varepsilon _{2}^{\prime }$
of the surrounding medium, the number density $n$ of electrons and the
conductivity $\sigma $ of the metal. The analytic expression are shown to
reproduce very well the experimental data for nano-spheres, nano-rods and
nano-plates, and can be used readily for estimating the resonant frequency
of plasmonic nanostructures as a function of their geometry, composition and
surrounding medium.

\textbf{Supplementary Information} is linked to the online version of the
paper.

\textbf{Competing Interests} The authors declare that they have no competing
financial interests.

\textbf{Acknowledgment} The work was support by grants W911NF-12-1-0072 and
W911NF-14-1-0051 (computation facility) from the Army Research Office.

\textbf{\ Correspondence} Correspondence and requests for materials should
be addressed to G. Lu. ~(email: ganglu@csun.edu).


\end{document}